\documentclass{aa}  
\usepackage{graphicx}
\usepackage{amsmath}	
\usepackage{xcolor}

\usepackage{txfonts}
\definecolor{verde}{rgb}{0.,0.56,0.}
\definecolor{lightblue}{rgb}{0.1,0.6,0.93}
\definecolor{mblue}{rgb}{0, 0.5, 0.815}

\usepackage[colorlinks]{hyperref}
\hypersetup{linkcolor=lightblue,citecolor=mblue,filecolor=cyan,urlcolor=lightblue}

\newcommand{\dg}{\textit{Nube}}

\newcommand{\noisechisel}{\texttt{NoiseChisel}}
\newcommand{\sextractor}{\texttt{SExtractor}}
\newcommand{\scamp}{\texttt{SCAMP}}
\newcommand{\swarp}{\texttt{SWarp}}
\newcommand{\hipercam}{HiPERCAM}

\begin{document} 

\title{An almost dark galaxy with the mass of the Small Magellanic Cloud}

\author{Mireia Montes \inst{1,2,3}
        \and 
        Ignacio Trujillo \inst{1,2} 
        \and
        Ananthan Karunakaran \inst{4,5}
        \and
        Ra\'ul Infante-Sainz \inst{6} 
        \and
        Kristine Spekkens \inst{7,8}
        \and
        Giulia Golini \inst{1,2}
        \and
        Michael Beasley \inst{9}
        \and 
        Maria Cebri\'an \inst{1,2}
        \and
        Nushkia Chamba \inst{10}
        \and
        Mauro D'Onofrio \inst{11}        
        \and 
        Lee Kelvin \inst{12}
        \and
        Javier Rom\'an \inst{13,1,2}
}

\institute{Instituto de Astrof\'{\i}sica de Canarias, c/ V\'{\i}a L\'actea s/n, E-38205 - La Laguna, Tenerife, Spain  \email{mireia.montes.quiles@gmail.com}
\and
Departamento de Astrof\'isica, Universidad de La Laguna, E-38205 - La Laguna, Tenerife, Spain
\and
Space Telescope Science Institute, 3700 San Martin Drive, Baltimore, MD 21218, USA
\and
Instituto de Astrof\'isica de Andaluc\'ia (CSIC), Glorieta de la Astronom\'ia, 18008 Granada, Spain
\and
Department of Astronomy \& Astrophysics, University of Toronto, 50 St. George Street, Toronto, ON M5S 3H4, Canada 
\and
Centro de Estudios de F\'isica del Cosmos de Arag\'on (CEFCA), Plaza San Juan, 1, E-44001, Teruel, Spain
\and
Department of Physics and Space Science, Royal Military College of Canada P.O. Box 17000, Station Forces Kingston, ON K7K 7B4, Canada
\and
Department of Physics, Engineering Physics and Astronomy, Queens University, Kingston, ON K7L 3N6, Canada
\and 
Centre for Astrophysics and Supercomputing, Swinburne University of Technology, Hawthorn, VIC 3122, Australia
\and
The Oskar Klein Centre, Department of Astronomy, Stockholm University, AlbaNova, SE-10691 Stockholm, Sweden
\and
Department of Physics and Astronomy, University of Padova, Vicolo Osservatorio 3, 35122 Padova, Italy
\and
Department of Astrophysical Sciences, Princeton University, Princeton, NJ 08544, USA
\and
Kapteyn Astronomical Institute, University of Groningen, Landleven 12, 9747 AD Groningen, The Netherlands
}

\date{}

 
\abstract
 {Almost Dark Galaxies are objects that have eluded detection by traditional surveys such as the Sloan Digital Sky Survey (SDSS). The low surface brightness of these galaxies ($\mu_r$(0)$>26$ mag/arcsec$^2$), and hence their low surface stellar mass density (a few solar masses per pc$^2$ or less), suggests that the energy density released by baryonic feedback mechanisms is inefficient in modifying the distribution of the dark matter halos they inhabit. For this reason, almost dark galaxies are particularly promising for probing the microphysical nature of dark matter.
In this paper, we present the serendipitous discovery of \dg{}, an almost dark galaxy with $<\mu_V>_e\sim 26.7$ mag/arcsec$^2$. The galaxy was identified using deep optical imaging from the IAC Stripe82 Legacy Project. Follow-up observations with the 100m Green Bank Telescope strongly suggest that the galaxy is at a distance of 107 Mpc. Ultra-deep multi-band observations with the 10.4m Gran Telescopio Canarias favour an age of $\sim10$ Gyr and a metallicity of [Fe/H]$\sim-1.1$. With a stellar mass of $\sim4\times10^8$ M$_{\odot}$ and a half-mass radius of $R_e=6.9$ kpc (corresponding to an effective surface density of $<\Sigma>_e\sim0.9$ M$_{\odot}$/pc$^2$), Nube is the most massive and extended object of its kind discovered so far. The galaxy is ten times fainter and has an effective radius three times larger than typical ultra-diffuse galaxies with similar stellar masses. Galaxies with comparable effective surface brightness within the Local Group have very low mass (tens of 10$^5$ M$_\odot$) and compact structures (effective radius $R_e<1$ kpc). Current cosmological simulations within the cold dark matter scenario, including baryonic feedback, do not reproduce the structural properties of Nube. However, its highly extended and flattened structure is consistent with a scenario where the dark matter particles are ultra-light axions with a mass of m$_B$=($0.8^{+0.4}_{-0.2}$)$\times10^{-23}$ eV.}


\maketitle
%

\section{Introduction}

Many cosmological observations at large scales suggest that dark matter can be well described as a cold and collisionless fluid \citep[see e.g.,][]{White1978,Blumenthal1984,Davis1985,Smoot1992}. Nonetheless, the predictions of this model at galactic scales have faced an increasing number of challenges, such as the ``cusp-core problem", the ``missing satellite problem" or the ``too-big-to-fail problem" \citep[see e.g., ][]{Boylan-Kolchin2011,Weinberg2015,DelPopolo2017}. Many of these problems can be mitigated by the effect of baryon feedback on the dark matter distribution \citep[see e.g., ][]{Davis1992,Governato2010,DiCintio2014}. However, if the number of stars or their spatial density is low enough, it would be difficult to argue that stellar feedback could be responsible for affecting the dark matter distribution, since there would not be enough energy to change the location of the dark matter \citep[see e.g.,][]{Penarrubia2012,Onorbe2015}. In parallel, as long as the direct detection of dark matter particles remains out of reach, other alternatives to the cold dark matter model have gained traction to solve the small scale challenges. These include the warm dark matter scenario \citep[see e.g.,][]{Sommer-Larsen2001,Bode2001}, self-interacting dark matter \citep[][]{Spergel2000} or fuzzy dark matter \citep[composed of ultra-light axions with masses in the 10$^{-23}$-10$^{-21}$ eV range; see e.g.,][]{Sin1994,Hu2000,Matos2001}. For these reasons, the search for objects with extremely low stellar surface densities (where the effect of baryonic feedback is not expected to be relevant) promises to probe the microphysical nature of dark matter, i.e., the properties of the dark matter particle.

In recent years, our ability to detect more and more diffuse galaxies with broadband imaging has increased considerably \citep[see e.g.,][]{Sandage1984, Conselice2003, vanDokkum2015, Roman2017, Lim2020, Tanoglidis2021, Trujillo2021, Marleau2021, LaMarca2022, Zaristky2023}, and it is expected to continue to increase with the arrival of very deep optical surveys \citep[see e.g., ][]{Ivezic2019}. Within this population of very faint galaxies, the so-called ``almost dark" galaxies are of particular interest. These faint galaxies are missed in the optical catalogues of wide field surveys such as the Sloan Digital Sky Survey 
\citep[SDSS,][]{Eisenstein2011}. Although there is no definition of the surface brightness of an almost dark galaxy, given the SDSS surface brightness limit (i.e., $\mu_r\sim26.5-27$ mag/arcsec$^2$; 3$\sigma$ in 10\arcsec$\times$10\arcsec\ boxes), galaxies with central surface brightness fainter than $\mu_r(0)\sim26$ mag/arcsec$^2$ are very difficult to detect in SDSS catalogues. Therefore, we can use such a value as a rough definition of what an almost dark galaxy should be. These galaxies represent less than 1\% of the galaxies found in blind HI surveys such as Arecibo Legacy Fast Arecibo L-band Feed Array \citep[ALFALFA, ][]{Giovanelli2005}, and have HI masses between 10$^7$ and 10$^9$ M$_\odot$. Since the definition of an almost dark galaxy depends on whether it is typically detected in the SDSS, its maximum stellar surface mass density is a function of its stellar population properties (age and metallicity). In a scenario where the age of the stellar population is old ($\sim10$ Gyr) and metal-poor ([Fe/H]$\sim-1$), a surface brightness of $\mu_r(0)\sim26$ mag/arcsec$^2$ corresponds to a few M$_\odot$/pc$^2$. This is very close to the expected stellar surface density resulting from the gas density threshold for star formation \citep{Schaye2004} and in good agreement with the values found at the edges of galaxies where star formation suddenly drops off \citep{Trujillo2020,Chamba2022}. Consequently, almost dark galaxies may also be an interesting place to study how galaxy formation occurs at low densities.

In this paper, we describe the structural properties of a very extended almost dark galaxy serendipitously discovered in the IAC Stripe82 Legacy Project \citep{Fliri2016, Roman2018_S82}. This object (which we have named \dg{}\footnote{$\backslash$noo-beh$\backslash$. Cloud in Spanish.}) was found during a visual inspection of one of the survey fields. The object is not visible in SDSS and appears quite noisy even in deeper images such as those produced by the Stripe82 data (see Fig. \ref{fig:fov}). Dedicated observations with the Gran Telescopio Canarias and the Green Bank Telescope have allowed to characterise its nature in detail. This paper presents a comprehensive analysis of the stellar populations and structural properties of this object. We also consider the properties of this very faint galaxy within the cold dark matter and fuzzy dark matter scenarios. 

Throughout this work we adopt a standard cosmological model with the following parameters: $H_0=70$ km s$^{-1}$ Mpc$^{-1}$, $\Omega_m=0.3$ and $\Omega_\Lambda=0.7$. Based on the probable redshift of this galaxy, the assumed distance is $107$ Mpc, corresponding to a spatial scale of $0.5039$ kpc/arcsec. All magnitudes in this paper are in the AB magnitude system.


\section{Data}

The data used in this paper come from two different facilities: the 10.4m Gran Telescopio Canarias (GTC) and the 100m Robert C.\ Byrd Green Bank Telescope (GBT). The details of each observation are described below.

\subsection{GTC \hipercam{} images}\label{sec:hipercam}

Deep optical multi-band imaging of \dg{} was performed using \hipercam{}.
\hipercam{} \citep{Dhillon2018} is a quintuple-beam, high-speed astronomical imager capable of simultaneously imaging celestial objects in five different Sloan filters ($u$, $g$, $r$, $i$, $z$). The image area of each of the five CCDs is $2048\times1024$ pixels ($2.7\arcmin \times1.4\arcmin$; 1 pixel$=0\farcs08$) divided into four channels of $1024\times512$ pixels each. \dg{} was observed on 9 and 10 January 2019. We followed the dithering strategy described in \citet{Trujillo2016} to reduce as much as possible the scattered light from the telescope structure.

\begin{figure*}
 \centering
   \includegraphics[width = 0.7\textwidth]{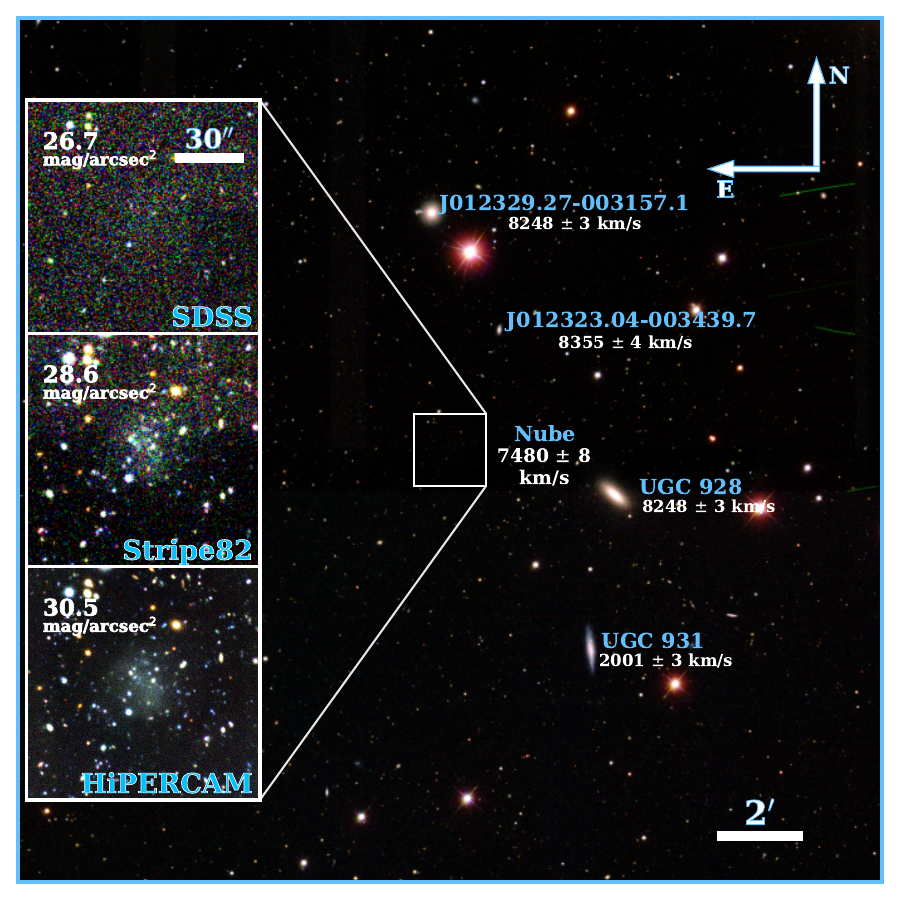}
   \caption{SDSS RGB image of the 20\arcmin$\times$20\arcmin\ region around \dg. The insets show a zoom in on the galaxy, showing how it appears on images with different limiting surface brightness in the r-band (26.7, 28.6 and 30.5 mag/arcsec$^2$; 3$\sigma$ on 10\arcsec$\times$10\arcsec boxes). Previously known galaxies in the field of view are labelled.
   \label{fig:fov}}
    \end{figure*}

The data reduction of the \dg{} images follows the steps detailed in \citet{Montes2020} and briefly described here. The entire reduction process was carried out in a controlled and enclosed environment as described in \citet{Akhlaghi2021}. After the standard calibration per CCD channel (bias and flat field), each set of four channels was assembled into a single image. The different exposures that went into the final images were visually inspected, and those with low quality (too noisy and/or very bright background) or strong gradients were discarded. Photometric calibration of these images was performed using SDSS DR12 \citep{Alam2015}. The size of \dg{} on the sky (effective diameter of $\sim$27$\arcsec$; see Fig. \ref{fig:fov}) is significantly smaller than the FOV of the camera ($162\arcsec\times 84\arcsec$), allowing reliable background subtraction and study of the galaxy using these images. The background subtraction was performed by subtracting a calculated constant value from the masked image. In addition to masking the foreground and background sources, we masked a  circular region with a radius of $40\arcsec$ centred on \dg{} (that is $\sim3$R$_\mathrm{e}$ of this galaxy, see Sec. \ref{sec:prof}) to avoid over-subtracting the light associated with the outer parts of the galaxy.

The final exposure time \textit{on-source} is 1 hour and 8 minutes for each band. The limiting surface brightness depths of the final images measured in $10\arcsec\times10\arcsec$ boxes are 30.5, 31, 30.5, 30 and 29.2 mag/arcsec$^2$ for $u$, $g$, $r$, $i$ and $z$ ($3\sigma$ above the background) respectively, measured using the method described in Appendix A of \citet{Roman2020}. The limiting magnitudes for point-like sources are: 26.2, 26.7, 26.2, 25.8 and 25 mag from u to z (5$\sigma$ within an aperture of 2\arcsec).

\subsection{GBT HI data}\label{sec:hIdata}

We performed $\sim$12 hours of observations with the Green Bank Telescope (GBT)\footnote{The Green Bank Observatory is a facility of the National Science Foundation operated under cooperative agreement by Associated Universities, Inc.} along the line of sight (LOS) to \dg{} between January 2019 and July 2019 (programmes GBT18B-356 and GBT19A-485). Our observational configuration is identical to that of \citet{Karunakaran2020b}, where we used the L-band receiver and the VErsatile GBT Astronomical Spectrometer (VEGAS) in Mode 7 (spectral resolution = 3 kHz $\sim0. 7$ km/s, bandpass = 100 MHz).\ The use of such a wide bandpass, sensitive out to recession velocities of $\sim14 000$ km/s, is essential in the blind search for HI in such a faint object, which can be at any distance along the LOS. 

\begin{figure}
 \centering
   \includegraphics[height = 0.26\textheight]{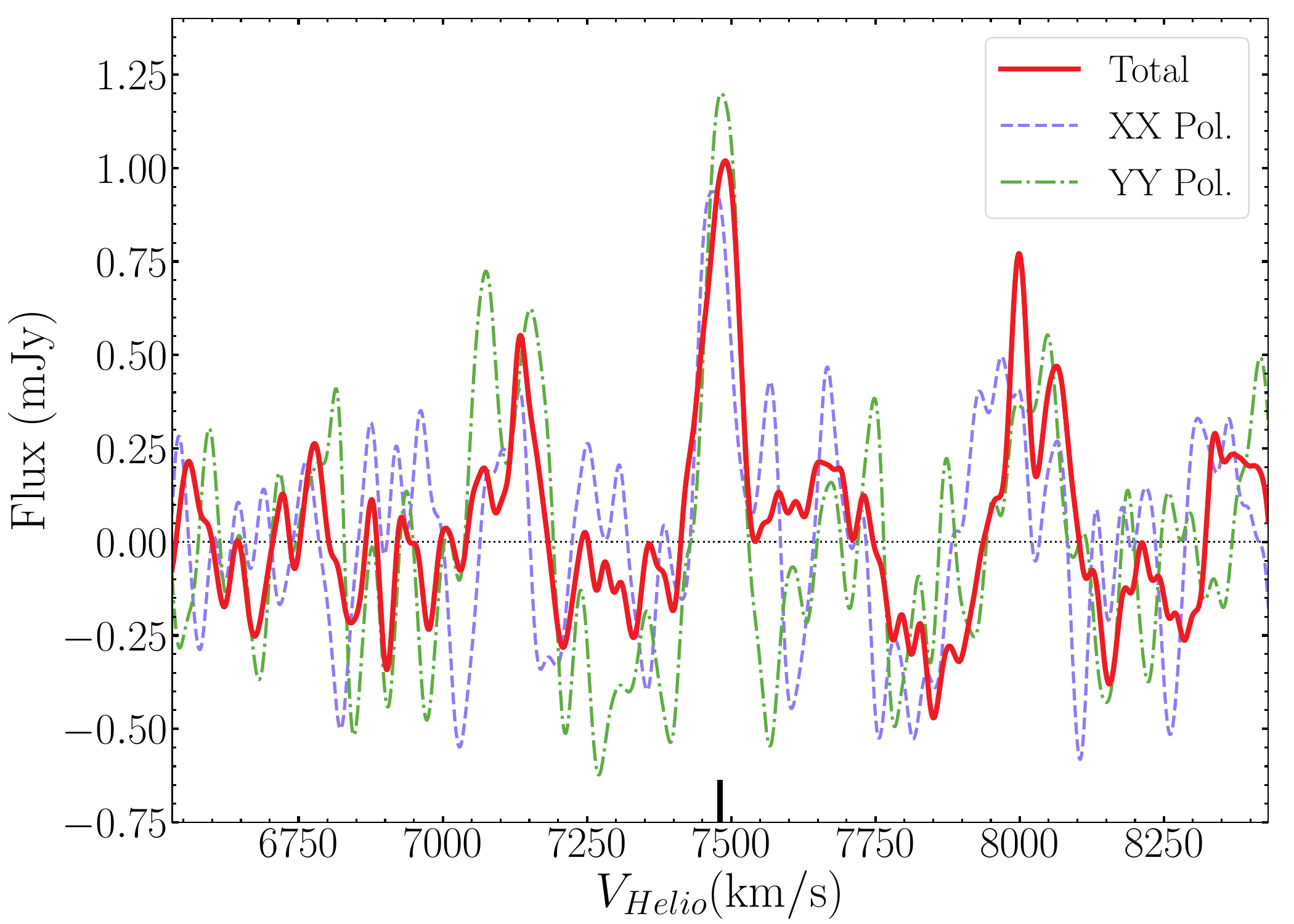}
   \caption{HI detection along the line of sight to \dg. We also show different polarisations of the data, XX and YY, to better identify which other HI emission peaks in the spectrum are potentially spurious. The black tick indicates the location of the peak with the strongest signal in our data. Note that other distinct peaks are not always present in the two different polarisations, thus reducing their relevance as real signals.
   \label{fig:hi}}
    \end{figure}

These data were reduced using the standard GBTIDL\footnote{\url{http://gbtidl.nrao.edu/}} procedure \textit{getps}.\ We removed both narrowband and broadband Radio Frequency Interference (RFI) before searching for potential HI signals. We note that broadband RFI (i.e.\ GPS L3, $\mu1.381$ GHz) was particularly prominent, resulting in almost a third of the data being flagged. The effective integration time of the final calibrated spectrum is therefore $\sim$8.5 hours. We also note that the fluxes are scaled up by a factor of 20\% due to the systematic offset in the GBT noise diode calibration values \citep{Goddy2020}. We smoothed the resulting RFI-free spectrum at various spectral resolutions to search for possible HI emission. Fig. \ref{fig:hi} shows a faint HI detection at V$_{Helio}$ = $7480 \pm 8$ km/s (red line) with a signal-to-noise ratio of 6.1, along with the statistically independent XX and YY polarizations (blue and green lines, respectively) that were co-added to produce the total spectrum. Although faint, the line is detected in both XX and YY, and it is also consistently detected in jackknifes about other observables such as observing date and time. For a number of reasons, which we explain below, we believe that this detection is most likely associated with \dg.

We estimate the systemic heliocentric velocity ($V_{Helio}$) and velocity width ($W_{50}$) of the HI detection using the methods described in \citet{Karunakaran2020a, Karunakaran2020b}, which are based on those of \citet{Springob2005}. Briefly, we fit first order polynomials to each edge of the HI profile between 15\% and 85\% of the peak flux and find the velocities corresponding to the 50\% flux value. The mean and difference of these velocities give $V_{Helio}$ and $W_{50}$ respectively.\ We correct $W_{50}$ for instrumental and cosmological redshift broadening according to \citet{Springob2005} and for ISM turbulence according to \citet{VerheijenSancisi2001} (see their section 4; see also \citealt{Karunakaran2020b}).

We convert $V_{Helio}$ to a kinematic distance ($D$) using the Hubble-Lema\^{i}tre law and assuming an uncertainty of 5 Mpc due to peculiar velocities. The derived distance is $107 \pm 5$ Mpc. We measure the HI flux ($S_{HI}$) by integrating over the line profile whose uncertainties are dominated by the noise and the 2\% diode uncertainty \citep{vanZee1997}. The flux value is $S_{HI} = 0.083\pm0.02$ Jy km/s. We use the kinematic distance together with the integrated flux in the standard equation for an optically thin gas \citep{HaynesGiovanelli1984} to calculate the HI mass:

\begin{equation} \label{eq:himass} 
\mathrm{M}_{HI}=2.356\times 10^{5}D^{2}S_{HI} \, [M_\odot]
\end{equation}
resulting in $\mathrm{M}_{HI} = 2.2^{+0.7}_{-0.5} \times10^{8} M_\odot$. 
All derived properties from the HI spectrum are at a spectral resolution of $25$ km/s and are listed in Table \ref{table:dgs}.\ 

\subsection{Possible sources of contamination in HI}\label{sec:hIcont}

Although it is tempting to associate the HI detection with the \dg{} galaxy, it could be that the measurement corresponds to another galaxy in the line of sight. For this reason, a detailed analysis of the velocity distributions of the galaxies in the \dg{} region is needed to see if the HI detection is compatible with any of them.

\begin{figure*}
 \centering
   \includegraphics[width = \textwidth]{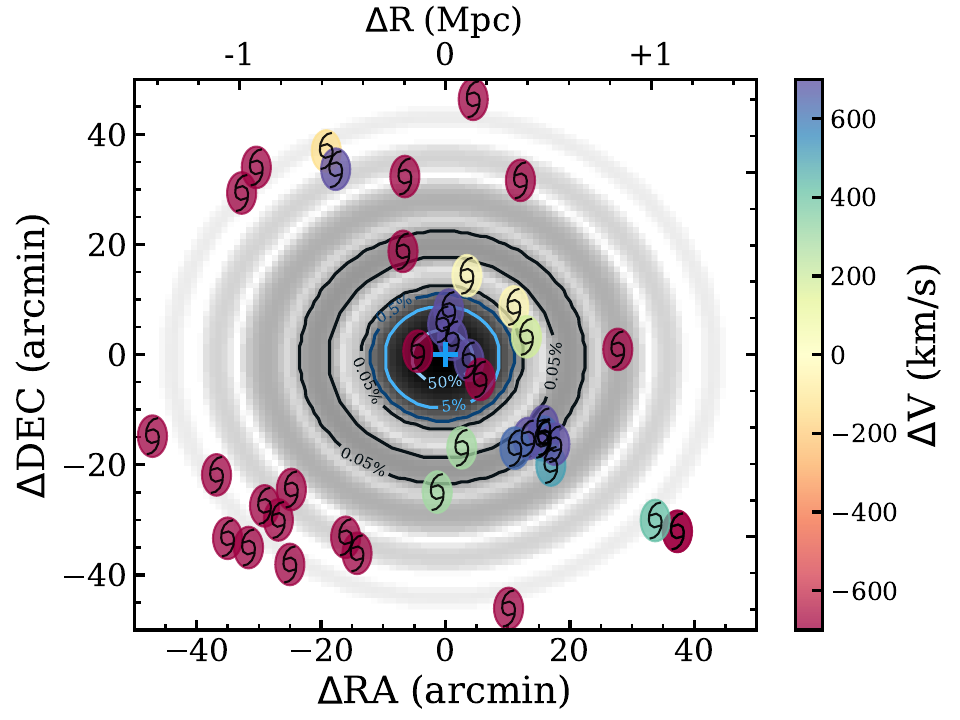}
   \caption{Galaxies found in NED with known spectroscopic redshifts in a range of $100\arcmin \times 100\arcmin$ around \dg{} and a recession velocity within 2000 km/s of that of the HI detection (i.e. $V_{helio} = 7480$ km/s). The galaxies are colour-coded according to their relative velocity with respect to the potential velocity of \dg{}. The grey-scale background image is the GBT beam response from \citet{Spekkens2013}. The contours indicate where the efficiency drops by the factor indicated on the label, from 0.05\% (black) to 50\% (light blue). The top horizontal axis indicates the equivalent size in Mpc at a distance of 107 Mpc. 
   \label{fig:galaxies}}
\end{figure*}

Using NED\footnote{The NASA/IPAC Extragalactic Database (NED) is operated by the Jet Propulsion Laboratory, California Institute of Technology, under contract with the National Aeronautics and Space Administration} we have created a catalogue of all objects in this database with spectroscopic redshifts in a region centred around \dg{} of $100\arcmin\times100\arcmin$. The total number of galaxies found is 711. All galaxies with recession velocities in the range 5480$<$$V_{helio}$$<$9480 km/s, 42 galaxies, are plotted in Fig. \ref{fig:galaxies}. The galaxies are colour coded according to the relative velocity of the HI detection. In this region, there are many galaxies with relative velocities $\Delta V$$<$-600 km/s from the HI detection (red symbols in the figure). There is also a smaller group with velocities about 500 km/s greater than the HI detection (purple symbols close to the centre). Interestingly, there is no galaxy with known spectroscopic redshift within a radius of 10\arcmin\ centred on \dg{} that has a $|\Delta V| < 500$ km/s. This is important because this is the region where potential contamination from a gas-rich source could mimic the signal we identify as possibly coming from \dg{}. To better illustrate this point, in Fig. \ref{fig:galaxies} we also show the 2D GBT beam response using the well-characterised pattern at 1.4 GHz from \citet{Spekkens2013}. The contours (from blue to black) indicate where the beam efficiency drops by the factor indicated on the label.

The closest galaxy in terms of velocity ($\Delta V = -17.8$ km/s) would be the spiral galaxy to the north: UGC929, in yellow in Fig. \ref{fig:galaxies}. This galaxy is 14.7 arcmin or 435 kpc away at the distance corresponding to the HI detection (i.e., 107 Mpc). This object is close to the first minimum of the beam response, shown by the grey-scale background image in Fig. \ref{fig:galaxies}, which means that the HI emission, if coming, from UGC929 would be suppressed by a factor of $\sim$$10^4$. If the source of the HI emission is UGC929, this implies that the HI mass of this galaxy should be $\sim$$10^{12}$ M$_\odot$. Using scaling relations between stellar mass and HI mass, a galaxy with the stellar mass of UGC 929 ($M_* = 2.4\times10^{10}M_\odot$, see Appendix \ref{app:ugc929}) is expected to have a mass in HI of $\sim2\times10^9M_\odot$ and no more than $\sim2\times10^{10}M_\odot$ \citep{Feldmann2020}. This is at least 100 times less gas than what it is needed to reproduce the observations. It is, therefore, very unlikely that the emission we see at the location of \dg{} is caused by the contamination from UGC929. To the northwest in Fig. \ref{fig:galaxies}, there is another galaxy at a similar velocity ($\Delta V = -27.8$ km/s) at 14\arcmin\ away. This galaxy appears to be a low-mass galaxy, probably a satellite of UGC929. Given it is also located in the first minimum of the beam response and its low mass, it is also very unlikely that it is the source of the HI emission.

None of the galaxies with a confirmed spectroscopic redshift appear to be responsible for the detection of HI in the \dg{} region. However, it is possible that the HI emission is coming from a low-mass galaxy in the region which, due to its faintness, has no spectroscopic redshift. To investigate this, we have made a catalogue of all sources with a photometric redshift within a radius of 10\arcmin\ around \dg{}. This radius corresponds to the area where the response efficiency of the GBT beam is greater than 5\% (see Fig. \ref{fig:galaxies}). We used one of the value added catalogues from the DR9 DECaLS survey, which provides photometric redshifts of sources down to a $5\sigma$ depth of $m_r$ = 24 mag \citep{Zhou2021}. The list of objects with a photometric redshift compatible with the HI emission (i.e., z=0.02502) is given in Table \ref{app:photz}. The vast majority of objects with a photometric redshift compatible with the HI detection reported in this work are either a star or a point-like source. This is somewhat to be expected, since at a given magnitude photometric catalogues are biased towards objects with higher signal-to-noise, and therefore extended objects are less represented or missing.

Finally, to account for the possibility that some faint and diffuse galaxies might not have a photometric redshift measured, we visually inspected the inner 10\arcmin\ around \dg{} using the Dark Energy Camera Legacy Survey \citep[DECaLS,][]{Dey2019} DR10 images to find extended faint sources that could potentially be emitting in HI. We found only two very faint galaxies, which are listed in Table \ref{table:other_cont}. These two galaxies are visually smaller than \dg{} and therefore potentially less massive at the same distance. Furthermore, these objects are found at a radial distance of $\gtrsim$ 4.5\arcmin\ from the central direction of the HI beam. While we cannot completely rule out the possibility that one of these galaxies is responsible for the HI detection, their off-centre location would imply that they are less likely to be associated with the HI detection than \dg{}.

The analysis in this section reinforces the idea that the detection of an emission peak in HI could be related to the optical detection of \dg{}. Moreover, the fact that the HI emission observed coincides with the velocity of UGC929 (even if it is extremely unlikely to be caused by it) is a further argument in favour of the HI detection being genuine and not a fluctuation. Having ruled out contamination of nearby HI sources, there is still the possibility that the optical and HI detections are unrelated, and therefore the distance assumed for the optical counterpart of the galaxy may be incorrect. However, this possibility does come with its own problems. For a detailed discussion of the implications of the distance for \dg{}, see the Appendix \ref{app:distance}.
The scenario that follows from the analysis in this section is that \dg{} could be a satellite of UGC929, at a distance of 107 Mpc.

\section{Analysis and results}

\subsection{Radial profiles of \dg}\label{sec:prof}

The aim of this paper is to study the properties of \dg{} in detail in order to determine its origin. A fundamental way to characterise the shape of a galaxy is through its radial surface brightness profiles. In addition, multi-wavelength information provides valuable constraints on galaxy formation processes. We have, therefore, derived the surface brightness radial profiles of this galaxy using the optical, multi-wavelength, \hipercam{} data. 

To measure the photometry of this galaxy, we first masked out all foreground and background sources in the image. As can be seen in the insets in Fig. \ref{fig:fov}, the galaxy is very faint and almost transparent. Since we cannot determine whether the clumps on top of the galaxy are part of the galaxy or are objects in the line of sight of the galaxy, we masked them all, leaving only the diffuse component for photometry. We used a combination of \noisechisel{} \citep{Akhlaghi2015, Akhlaghi2019} and manual masking after visual inspection. This is a very conservative approach as we are only analysing the diffuse light of the galaxy. Appendix \ref{app:mask} shows the mask used in this work.

\begin{figure}
 \centering
   \includegraphics[width = 0.47\textwidth]{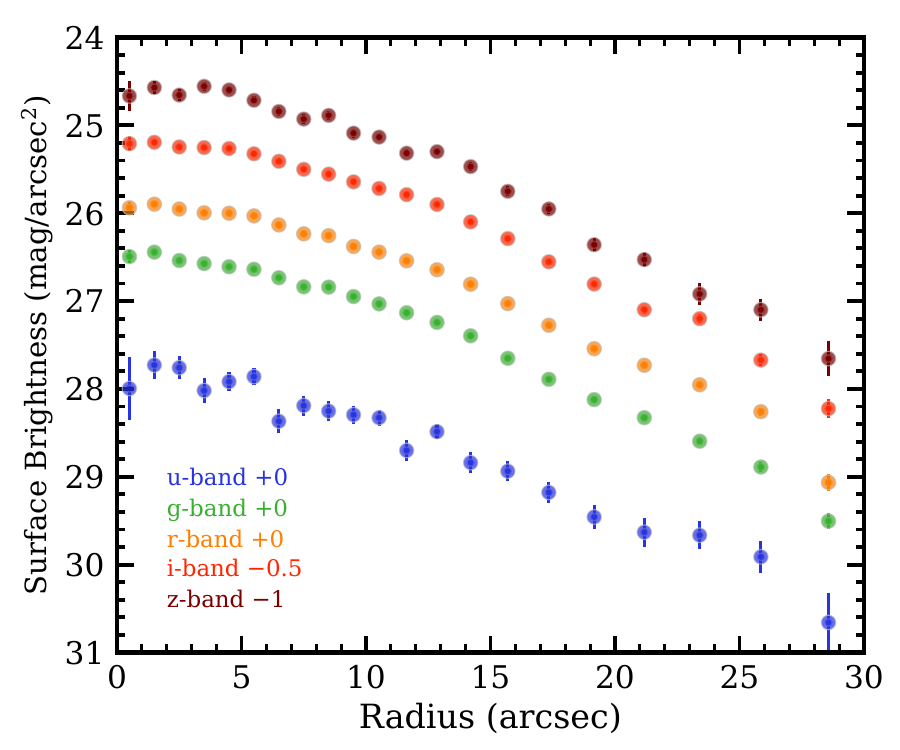}
   \caption{Radial surface brightness profiles of \dg{} in the Sloan u, g, r, i and z bands. The profiles in the i and z bands have been shifted vertically for ease of visualization.
   \label{fig:profiles}}
    \end{figure}

Once we had masked out all the sources of contamination, we derived the radial surface brightness profiles of the galaxy. Given the nearly circular shape of the object, we decided to extract the radial profiles using circular annuli at different radial distances, up to 30\arcsec\ from the centre of the galaxy. To derive the profiles, we use a  custom \texttt{python} code. For each radial bin, the surface brightness was obtained as the $3\sigma$-clipped median of the pixel values. Fig. \ref{fig:profiles} shows the surface brightness profiles of \dg{} for the 5 bands imaged with \hipercam. Some of the profiles are shifted by a constant value, given in the legend, for ease of viewing. The errors are calculated as a combination of the Poisson noise in each annulus and the error in the sky given by the distribution of background pixels in each image. These surface brightness profiles are corrected for the absorption of our Galaxy \citep[E(B$-$V) = 0.34,][]{Schlafly2011}.

 \begin{figure}
 \centering
   \includegraphics[width = 0.47\textwidth]{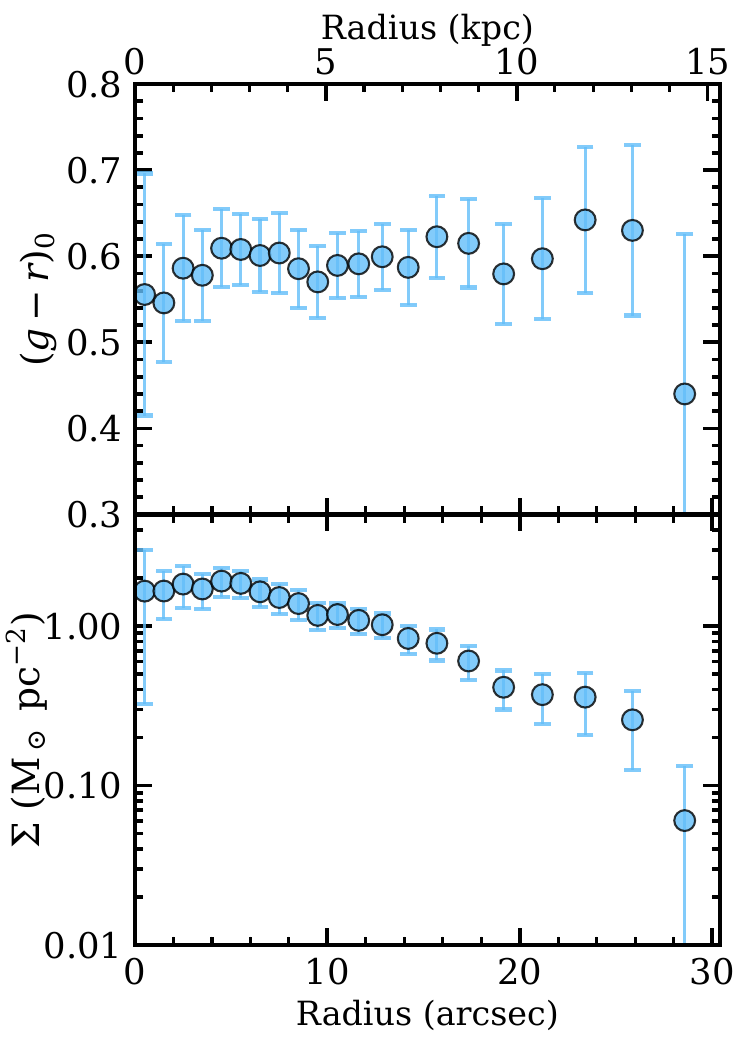}
   \caption{Upper panel: the \hipercam{} $g-r$ colour profile of \dg. Bottom panel: Stellar surface mass density profile of \dg. The upper horizontal axis indicates the equivalent size in kpc at a distance of 107 Mpc. \label{fig:color}}
    \end{figure}

We can now study the radial variations of the galaxy's stellar populations. The surface brightness profiles are used to determine the radial $g-r$ profile of the galaxy. The top panel in Figure \ref{fig:color} shows the $g-r$ colour profile of \dg. The colour profile appears flat at all radii, with an average colour of $\sim$0.6. This almost flat colour profile could be compatible with a similar age and metallicity at all radii, up to $10\arcsec$ ($15$ kpc). 

The bottom panel of Fig. \ref{fig:color} shows the surface stellar mass density profile for this galaxy. To derive it, we follow the procedure given in \citet[][]{Bakos2008} to link the observed surface brightness in the g-band to the radial variation of the stellar mass to light (M/L) ratio. The M/L ratio was obtained from the prescriptions given in \citet{Roediger2015} assuming a \citet{Chabrier2003} IMF, using the $g-r$ colour profile. As with the colour profile, the stellar mass density profile of \dg{} is also relatively flat compared to the stellar mass density profiles of galaxies with similar stellar masses \citep[e.g.,][]{Montes2021b}. We expand on this result later in the text.

We have used this stellar mass density profile to measure the radius that encloses half the mass of the galaxy, or the half-mass radius ($R_{\mathrm{e}}$). This radius is $R_{\mathrm{e}} = 13.7\pm1.7$ arcsec ($6.9\pm 0.8$ kpc at a distance of 107 Mpc). The effective surface stellar mass density of \dg{} is $<\Sigma>_{\mathrm{e}}= 0.9\pm0.1$ M$_{\odot}$ pc$^{-2}$. The total stellar mass of the galaxy derived from the surface mass density profile, assuming circular symmetry, is M$_* = 3.9\pm1.0\times 10^8$ M$_\odot$. The effective surface brightness is $<\mu_V>_e = 26.75\pm0.02$ mag/arcsec$^2$. These values, along with other global properties of \dg, are listed in the Table \ref{table:dgs}.

To assess the validity of our results, we also fitted  \citet{Sersic1968}  models \citep[using the code GALFIT][]{Peng2002} to the galaxy in all the different \hipercam{} bands, independently. The values of the half-light effective radius, $r_e$, and the S\'ersic index for each band are given in Table \ref{table:galfit}. The values of $r_e$ from the GALFIT fits are consistent with those derived using the surface brightness profiles.

\subsection{Stellar populations of \dg}\label{sec:sed}

\begin{figure}
 \centering
   \includegraphics[width = 0.42\textwidth]{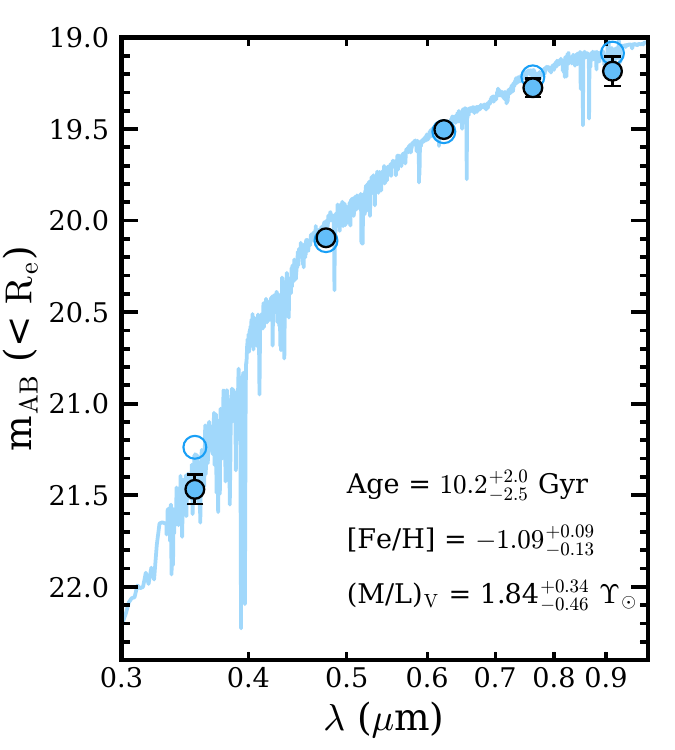}
   \caption{Spectral energy distribution (SED, filled circles) of \dg{} derived from \hipercam{} photometry within the half-mass radius. The best-fit single stellar population model is shown as the blue line and the best-fit magnitudes (model convolved with the filter response) are the open circles. The age, metallicity and mass-to-light ratio corresponding to the best fit model are given in the figure. \label{fig:fit_sed}}
    \end{figure}

Given the very low surface brightness of this galaxy ($\mu_V(0)$$\sim$26.2 mag/arcsec$^2$), the use of deep multi-wavelength  observations is the best, if not the only, way to constrain its stellar populations, namely the age, metallicity and stellar mass-to-light ratio (M/L). To estimate these quantities, we first constructed the spectral energy distribution (SED) of the galaxy (see Fig. \ref{fig:fit_sed}). The photometry at each band was derived measuring the total flux within a circular aperture with radius equal to the half-mass radius $R_e$ (13.7\arcsec) of the galaxy. The purpose of using this aperture (instead of the total galaxy) is to ensure enough signal to reliably characterise the SED of \dg{}. The errors of the photometry of each individual band that compose the SED are a combination of the photometric errors and the zero point uncertainties.

To characterise the stellar population properties of the galaxy, we fitted \citet{Bruzual2003} single stellar population (SSP, instantaneous burst) models to the SED of the galaxy. 
This is a reasonable assumption given the homogeneous colour distribution of this galaxy (Fig. \ref{fig:color}). 
The parameters to be fitted are three: age, metallicity and luminosity, and we assumed a Chabrier Initial Mass Function \citep[IMF,][]{Chabrier2003} for the models. 
Since the SED is derived from broadband imaging, it is a good approach to assume an SSP to describe the stellar populations of this galaxy, since the information we have is limited. This assumption will give us the average properties of this galaxy. For the fit, we use the $\chi^2$ minimisation approach described in \citet{Montes2014}. We derive a most likely age of $10.2^{+2.0}_{-2.5}$ Gyr and a metallicity [Fe/H] of $-1.09^{+0.09}_{-0.13}$, values in agreement with those of old diffuse low-mass galaxies \citep[see e.g.,][]{Ruiz-Lara2018, Heesters2023, Iodice2023}. Uncertainties in the parameters have been estimated by marginalising the 1D probability distribution functions obtained during the fitting. Since the colour profile of this galaxy is flat at all radii (see Fig. \ref{fig:color}), we assume that the stellar populations obtained for the inner $R_e$ are representative for the whole galaxy. 

Using the M/L derived from the SED fit ($(\mathrm{M/L})_\mathrm{V}= 1.8^{+0.3}_{-0.5}$)\footnote{This M/L value is consistent with the M/L ratios derived at each radial distance using the $g-r$ colour profile in Sec. \ref{sec:prof}; a median of $1.7$.}, the total stellar mass of \dg{} is M$_* = 4.4\pm0.8 \times10^8$ M$_\odot$, in agreement with the stellar mass estimate derived in Sec. \ref{sec:prof}. As this stellar mass is derived with the properties inside $R_e$, we prefer to use the previous estimate (Sec. \ref{sec:prof}) as it is more representative of the whole galaxy.

\begin{table*}[t]
\centering
\tabcolsep=0.2cm
\begin{tabular}{cccccccc}
\hline \hline
RA &  DEC  & z & Distance & log(M$_{\mathrm{HI}}$/M$_\odot$) &  log(M$_*$/M$_\odot$) & $R_{\mathrm{e}}$ & $\mu_V(0)$\\
& &  & (Mpc) &   &  & (arcsec) & (mag/arcsec$^2$) \\
\hline
01$^h$23$^m$27.37$^s$ & -00$^d$37$\arcmin$27.83$\arcsec$ & 0.02502$\pm0.00003$  &  $107 \pm 5$ &  $8.35 \pm 0.12$ &  $8.6\pm 0.1$& $13.7\pm 1.7$ & $26.23\pm0.07$ \\
\hline \hline
$W_{50}$& $<\mu_V>_e$  & $<\Sigma>_e$ & Age & [Fe/H] & $b/a$ & $R_e$\\
km/s & (mag/arcsec$^2$)  & (M$_{\odot}$pc$^{-2}$) & (Gyr) & & & (kpc)\\
\hline
 $34 \pm11$ &$26.75 \pm 0.02$  & $0.9\pm0.1$  & $10.2^{+2.0}_{-2.5}$ & $-1.09^{+0.09}_{-0.13}$ & 0.97$\pm$0.01 & $6.9\pm0.8$\\
\hline
\end{tabular}
\caption{Global properties of \dg. The distance-dependent parameters have been calculated assuming that the object is at a distance of 107 Mpc. }
\label{table:dgs}
\end{table*}

\subsection{Dynamical mass of \dg}\label{sec:dynmass}
A rough estimate of the dynamical mass of this galaxy\footnote{This calculation assumes that \dg{} is in dynamical equilibrium. This is a reasonable assumption given the large distance to its potential progenitor UGC929.} can be obtained using the following equation from \citet{Spekkens2018}: 

\begin{equation}
    M^{3R_e}_{dyn} = 6.96\times 10^5 R_e \left(\frac{W_{50}}{2\times sin (i)} \right)^2\, [M_{\odot}]
\end{equation}

with R$_e$ the half-mass radius in kpc, $i$ the inclination of the HI disc in degrees, and W$_{50}$ the turbulence-corrected velocity width of the HI line. We calculate the inclination of the HI disc as follows:

\begin{equation}
    \cos(i)=(b/a)
\end{equation}

In Table \ref{table:dgs}, we provide the axis ratio, $b/a$, of \dg{} obtained from GALFIT S\'ersic fits to the object (see sec. \ref{sec:prof}). The resulting value (0.97$\pm$0.01) is the average from the axis ratio measured in the $g$, $r$, and $i$ bands. That results in an inclination value of 13$\pm$3 deg. It is worth emphasizing that using the axis ratio of the stellar distribution to characterise the inclination of the HI disc is a rough approximation. The dynamical mass we derive for \dg{} is $2.6\pm 1.7 \times10^{10}$ M$_{\odot}$ within $3R_e = 20.7$ kpc.

\subsection{A search for Globular Clusters}\label{sec:gcs}

Globular clusters (GCs) provide a complementary way to estimate the distance to an extragalactic object based on the universality of their luminosity function. The peak of the GC distribution lies at M$_V = -7.6$ mag for different galaxies \citep[see e.g.,][]{Rejkuba2012}, making it a secondary distance indicator. Therefore, the detection of GCs in \dg{} would help us to validate the distance obtained with the HI detection. Note that the peak of the GC luminosity at the distance of 107 Mpc (i.e., a distance modulus of 35.15 mag) is expected at m$_g\simeq 27.5$ mag, fainter than our point-like 5$\sigma$ limiting magnitudes meaning that it would be difficult to detect GCs, if any, in our images (see section \ref{sec:hipercam}). We expect that an ultra-diffuse galaxy of similar stellar mass than \dg{} would have around $\sim$10 GCs, although the range varies from 0 to 30 GCs. However, only the brightest GCs of the system are expected to be detectable. 

We follow a similar procedure to detect globular clusters as in \citet{Montes2020, Montes2021b}. In short, we first preselect the GCs based on their morphology (size and ellipticity), and then we refine the selection based on their colours. However, in this case we do not have the high spatial resolution of the \emph{HST} images to make a preselection based on the morphology of the GCs, but only the ground-based images. For this reason, we have to impose very strict shape criteria to minimise contamination in our selection. 

We run \sextractor{} on our images in dual mode, with the $r$-band image as the detection image. At a distance of 107 Mpc, globular clusters will look like point sources in the images. Therefore, to pre-select the GC candidates, we first select stars in our images by imposing that the stellarity parameter in \sextractor{} (`CLASS\_STAR') is greater than 0.98. We do this to estimate the FWHM of a point-like source in our images, to impose strict conditions on our detections and to minimise contamination from other objects. We find 6 point-like objects, 3 of which were detected in Gaia DR3 \citep{GaiaCollaboration2022k}. The FWHM of these point-like objects are between $10$ and $14$ pixels (0.8\arcsec to 1.12\arcsec) and their ellipticities are $<0.1$. Therefore, based on these values, we pre-selected the GC candidates with: ellipticity $< 0.1$, $9 <$ FWHM $< 15$ pix and $m_r < 26.2$ mag, in order to minimise false detections.  
 \begin{figure}
 \centering
   \includegraphics[height = 0.35\textheight]{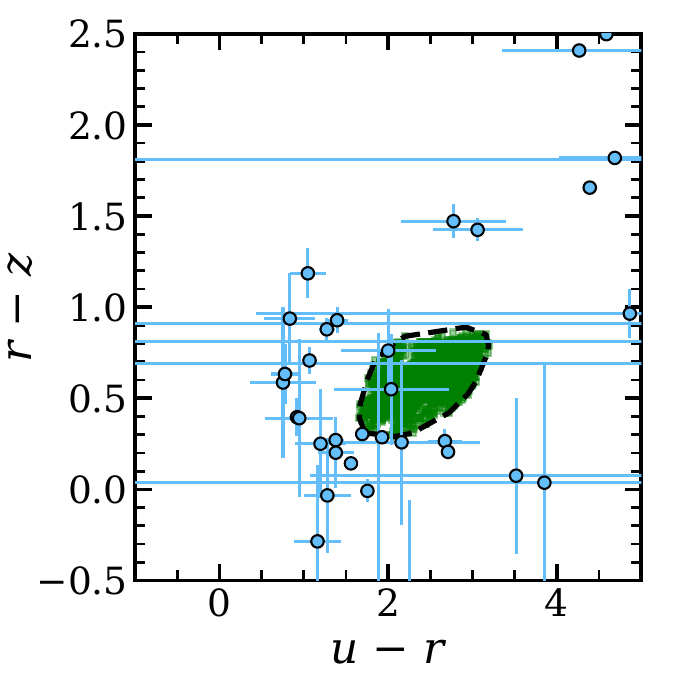}
   \caption{The $(u-r)-(r-z)$ colour-colour diagram of the initial sample of GC candidates (blue) of \dg. The green squares are the location of NGC5128 candidates in \citet{Taylor2017} with a probability greater than 95$\%$ of being true GCs. The black dashed line indicates the convex hull (envelope) computed as the smallest region containing all the green squares.
   \label{fig:gcs}}
  \end{figure}

After this initial selection, we also applied the $(u-r)-(r-z)$ colour-colour selection used in \citet{Taylor2017} to narrow down the selection of GCs\footnote{See \citet{Munoz2014, Lim2020, Cantiello2020} for other examples on using colour-colour diagrams to identify candidate GCs.}. To do this, we defined a colour-colour region based on the position on the diagram of candidate GCs with a probability of $>$0.95 of being true GCs in \citet{Taylor2017}, the green squares in Fig. \ref{fig:gcs}. To define this region, we computed the convex hull or convex envelope, i.e., the minimum region containing all GCs from \citet{Taylor2017} (dashed line in Fig. \ref{fig:gcs}). 

Only two sources fall on the area defined by the GCs provided by \citet{Taylor2017}\footnote{\citet{Taylor2017} use the GCs of NGC5128 to generate their catalogue. NGC5128 is a massive galaxy with intense merger activity. Therefore, the population of GCs in this massive galaxy includes GCs of a wide variety of origins, from those formed in-situ in this massive galaxy to those accreted by mergers with a large number of dwarfs. In this sense, the \citet{Taylor2017} catalogue is expected to contain a very heterogeneous and complete sample of GCs.}. However, both are very far from \dg{} ($>$110\arcsec or 54 kpc from its centre), so they are unlikely to be associated with the galaxy. We also checked the three sources near the bottom of the region (u-r$\sim$2, r-z$\sim$0.25), but they are also not spatially associated with \dg. In summary, we are unable to detect any GCs associated with \dg, as expected, considering both our point-like limiting magnitudes and if the distance to the galaxy is 107 Mpc.

\section{On the formation mechanism of \dg}\label{sec:discussion}

\subsection{\dg{} in comparison with other Low Surface Brightness galaxies}

\begin{figure*}
 \centering
   \includegraphics[height = 0.55\textheight]{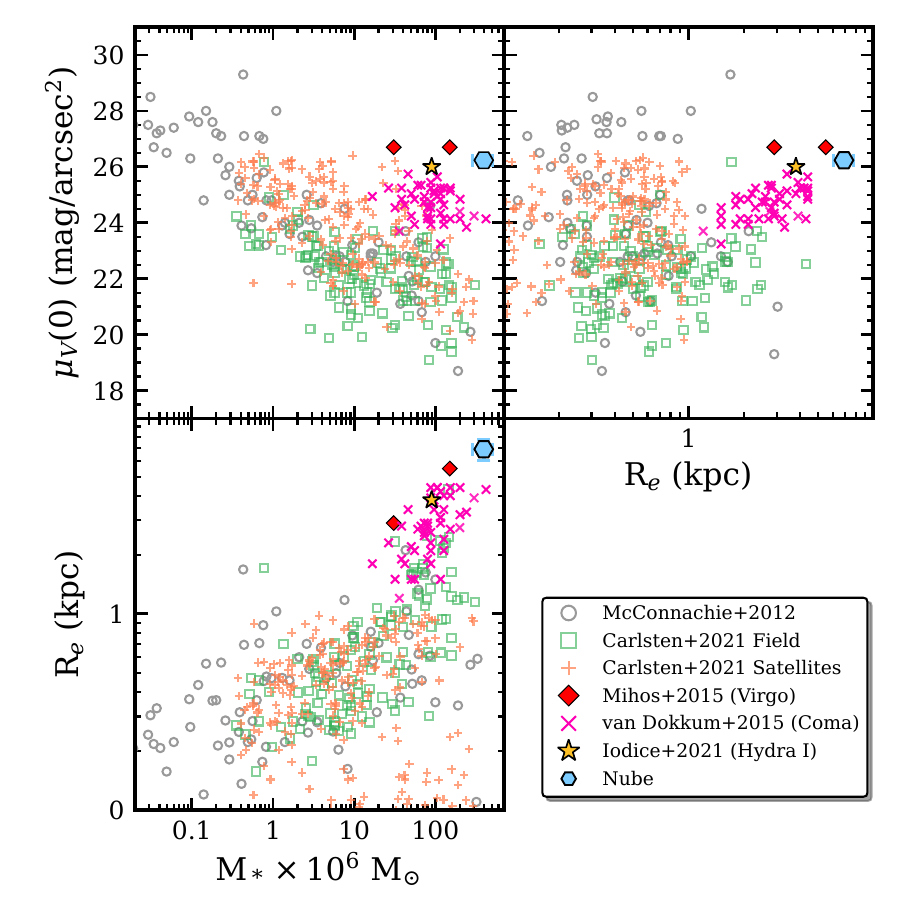}
   \caption{Comparison of the structural properties of \dg{} with respect to other Local Group galaxies from \citet[][ grey open circles]{McConnachie2012}, dwarf galaxies in \citet[][green open squares and orange crosses]{Carlsten2021}, UDGs in Virgo from \citet[][red diamonds]{Mihos2015}, UDG32 in Hydra I from \citet[][golden star]{Iodice2021} and Coma from \citet[][pink x]{vanDokkum2015}. Although the central surface brightness is typical of other dwarf galaxies, the stellar mass and half-mass radius make \dg{}, at a distance of 107 Mpc, an extreme object, even more so than the UDGs in Virgo and Coma.
   \label{fig:comparison}}
    \end{figure*}

The galaxy reported in this paper, \dg, has some extreme properties in terms of size and surface brightness. Fig. \ref{fig:comparison} shows the structural properties of \dg{} compared to other low surface brightness or dwarf galaxies compiled from the literature. The grey open circles represent low-mass Local Group galaxies from \citet{McConnachie2012}, the green open squares are field dwarfs, while the orange crosses are satellite dwarfs (orbiting a more massive galaxy) from \citet{Carlsten2021}, pink `x' are UDGs in Coma from \citet{vanDokkum2015}, the golden star is UDG32 from \citet{Iodice2021} and the red diamonds are two of the LSB galaxies in Virgo from \citet{Mihos2015}. Compared to all these objects, \dg{} appears quite unique due to its low central surface brightness and extension for a galaxy of its stellar mass. The closest objects in size, mass and central surface brightness are the two LSB galaxies in Virgo from \citet{Mihos2015}: VLSB-B and VLSB-C\footnote{To derive the mass of VLSB-C, we assume a B$-$V = 0.7 mag for this galaxy, which is typical for dwarf galaxies and the M/L colour relation in \citet{Bell2003}.} and UDG32 in Hydra I \citep{Iodice2021}. We do not include VLSB-A, as this LSB galaxy is clearly tidally perturbed, as indicated by its visible tidal tails. VLSB-C appears to be quite similar in size and brightness to \dg{} \citep[$R_e = 5.5$ kpc and $\mu_V(0) = 26.7$ mag/arcsec$^2$,][]{Mihos2015}. This galaxy shows no obvious signs of tidal stripping; possibly it is an LSB galaxy in the cluster outskirts or falling into the Virgo cluster for the first time. 

We also compare \dg{} with other galaxies that have been described as ``almost dark" in previous works. One example is AGC229385 \citep[also known as Coma P.;][]{Janowiecki2015, Brunker2019}. Unlike \dg, this galaxy appears elongated. It also has an extreme ratio of HI to stellar mass \citep[$M_{HI}/M_*$=81; ][]{Brunker2019} compared to \dg{} ($M_{HI}/M_* \sim 1$). This is also the case of AGC229101 \citep{Leisman2021}. This galaxy is an HI source ($M_{HI}/M_*$=98) with a very dim optical counterpart ($\mu_g(0) = 26.6$ mag/arcsec$^2$) and $R_e = 3$ kpc.  Other two extended ``almost dark" galaxies are AGC229398 and AGC333576 \citep{Gray2023}. Their effective radii are large, 9 and 4.4 kpc respectively, and their stellar masses ($\sim$10$^8$ M$_\odot$) are also close to \dg{}. However, their central surface brightness are significantly brighter than \dg{} (23.8 and 24.6 mag/arcsec$^2$). 

\begin{figure}
 \centering
   \includegraphics[width = 0.46\textwidth]{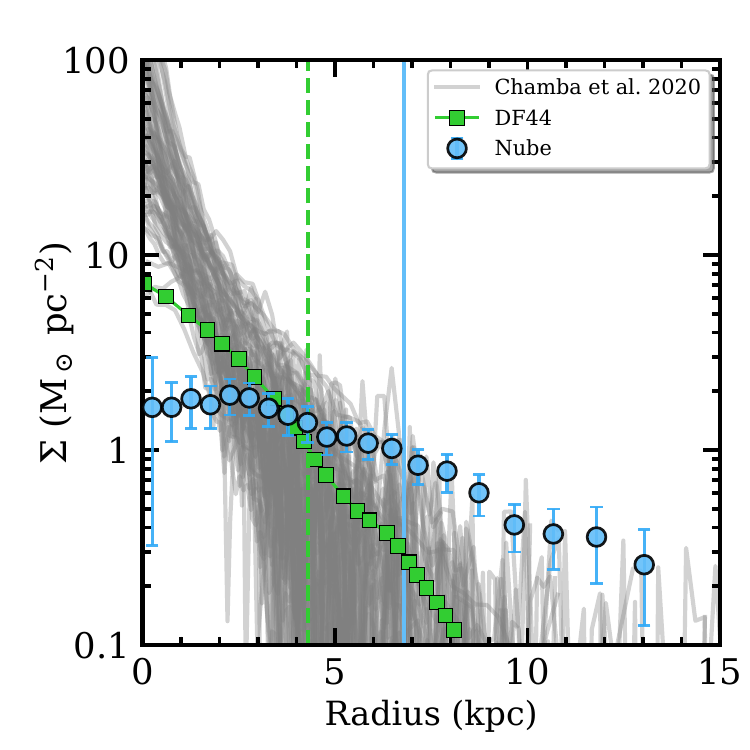}
   \caption{Surface stellar mass density profile of \dg{} (blue dots) compared with other galaxies of similar stellar mass. The profiles of dwarf galaxies from \citet{Chamba2020} in the same mass range (1-5$\times10^8 M_{\odot}$) have been plotted. We also show the profile of DF44 (green squares), as it is an iconic large ultra-diffuse galaxy. The vertical lines indicate the half-mass radius ($R_e$) for \dg{} (blue solid line) and DF44 (green dashed line). \label{fig:dens_comp}}
\end{figure}

In addition to comparing the global properties of \dg{} with other low-mass galaxies, it is also very instructive to compare its surface mass density profile with respect to galaxies of similar stellar mass. In Fig. \ref{fig:dens_comp}, we show the stellar surface mass density of \dg{} (blue dots) derived from the $g-r$ colours (Sec. \ref{sec:prof}). For comparison, we plot the profiles of dwarf galaxies of similar stellar mass (grey lines), in the mass range $1-5\times 10^{8}\, M_{\odot}$ from \citet{Chamba2020}. This shows how different \dg{} is from typical dwarf galaxies of similar mass. Even ultra-diffuse galaxies with large effective radius are not comparable to this galaxy. For example, the iconic ultra-diffuse galaxy Dragonfly 44 (DF44), also shown as the green squares in Fig. \ref{fig:dens_comp}, has a $r_e = 4. 3\pm0.2$ kpc \citep{vanDokkum2016}\footnote{Other estimates for $r_e$ for DF44 are $r_e = 3.3\pm0.3$ kpc \citep{Chamba2020} and $3.9\pm0.7$ kpc \citep{Saifollahi2021}.}, and a central surface brightness of $\mu_{g}(0) = 24.5$ mag/arcsec$^2$. This is 1.5 times smaller and $1.4$ mag/arcsec$^2$ brighter than \dg. 

\subsection{On the nature of \dg}\label{sec:born}

Given the extreme properties of \dg, it is interesting to discuss whether these properties are a result of the original formation of this galaxy, or whether they are due to a later evolutionary process caused by the environment in which it is found. In this section, we discuss two possible alternatives. In the first, we explore whether \dg{} can be considered a tidal dwarf galaxy. The second is whether \dg{} was born with a stellar density typical of dwarf galaxies of the same stellar mass and has been deformed by the environment into this peculiar structure. Either of these two possibilities (if correct) should be able to explain the stellar populations and the present morphology of \dg.

\subsubsection{Is \dg{} a Tidal Dwarf Galaxy (TDG)?}

There is a population of galaxies whose formation mechanisms would not be associated with what is expected to be the main channel of formation of galaxies, i.e., gravitational collapse of gas within a dark matter halo. These galaxies are known as Tidal Dwarf Galaxies (TDGs). They form from material torn away from larger galaxies by tidal interactions or by harassment \citep[e.g.,][]{Duc2008} and, as a result of their formation, they exhibit characteristics such as significant gas fractions, low dark matter content, and metallicity higher than expected for dwarf galaxies of similar mass \citep[e.g.,][]{Duc2012}.

Around 95\% of the known TDG galaxies are located at relatively modest distances ($\sim$20 kpc) from their potential progenitors \citep{Kaviraj2012}. However, \dg{} is very distant (435 kpc) from its likely progenitor UGC929. This large distance from its possible parent should not be considered as a factor against \dg{} being a TDG, since the selection of TDG galaxies usually requires a visible tidal tail and is therefore biased towards newly formed objects. Some simulations suggest that if the TDG is massive enough (M$>10^8 \,M_{\odot}$) and escapes the parent galaxy with sufficient velocity, it could become self-gravitating and avoid falling back into its progenitor system \citep{Bournaud2006}. However, it is unclear how long TDGs formed in this way could survive without the protection of a massive dark matter halo, as both axisymmetric and non-axisymmetric instabilities would eventually destroy the system \citep{Sellwood2022}. Assuming that these galaxies manage to survive, they are expected to remain gas-rich and very faint. These old galaxies are therefore good candidates to be identified as ``almost dark" galaxies \citep[see e.g.,][]{Cannon2015, Janowiecki2015, Leisman2017, Roman2021}.

The metallicity observed in \dg{} ([Fe/H] = $-1.09^{+0.09}_{-0.13}$) is similar to that of normal dwarf galaxies \citep[e.g.,][]{McConnachie2012}. However, it is lower than expected for currently forming TDGs \citep[solar metallicity, e.g.,][]{Duc2000}, i.e., more metallic as it inherits the metallicity of the more massive parent galaxies. Nonetheless, given the age of the galaxy (9 Gyr), it could have formed from ancient spirals with less enriched gas than the present ones \citep[see e.g.,][]{Recchi2015}. Therefore, the relatively low metallicity of \dg{} should not necessarily be an indication that it did not form as a TDG. 

Another characteristic of TDGs is their low dark matter content. The dynamical mass of \dg{} is $2.6\pm 1.7 \times10^{10}$ M$_{\odot}$ within $3R_e = 20.7$ kpc (Sec. \ref{sec:dynmass}), while the stellar mass is M$_* = 3.9\pm1.0 \times10^8$ M$_\odot$ (Sec. \ref{sec:sed}). This means that the ratio of dark matter to stellar mass is between $20 - 150$. This amount of dark matter is significantly larger than the amount of dark matter expected if the galaxy had formed as a TDG \citep[M$_{dyn}/M_{*}<$2; ][]{Gray2023}. In addition, its gas content relative to the stellar content is not very large compared to those found in other TDGs \citep[M$_{HI}/M_{*}\sim$10, ][]{Gray2023}. For these reasons, we conclude that \dg{} is with high confidence not a TDG.

\subsubsection{\dg{} as a result of environmental processes}

If \dg{} is not a TDG, can its structural properties be explained as the result of interaction with its environment? It is known that dwarf galaxies can be dynamically heated and ``puffed up'' by interactions with more massive galaxies \citep[e.g.,][]{Liao2019, Tremmel2020}. This puffing could explain why the age and metallicity of \dg{} are similar to other regular dwarf galaxies, while its physical properties (effective radius and central surface brightness) are extreme. However, against this scenario for the origin of \dg{} is the fact that the galaxy does not inhabit a particularly dense environment. In fact, its nearest massive neighbour, UGC 929, is at a distance of 435 kpc in projection. 

\begin{figure}
 \centering
   \includegraphics[width = 0.43\textwidth]{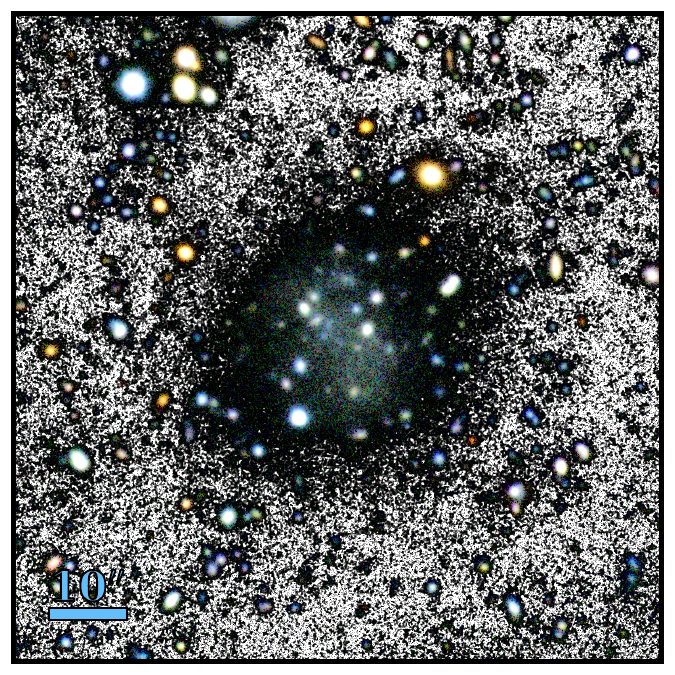}
   \caption{A region of $100\arcsec \times 100\arcsec$ around \dg. The figure is a composite of an RGB colour image using the $g$, $r$ and $i$ \hipercam{} bands and a black and white $g + r$ image for the background.
   \label{fig:hc_rgb}}
    \end{figure}

Assuming that \dg{} and UGC 929 have had a close encounter in the past, we can make a rough estimate of when this occurred. Considering the current projected distance between the two objects plus a transverse projected velocity of $\sim$100 km/s (i.e., 100 kpc/Gyr), the close encounter would have occurred 4 to 5 Gyr ago. Given the dynamical time of each galaxy, this is enough time to have erased any signature of the perturbation in the central regions of both objects. We do not find the central region of \dg{} to be obviously perturbed (see Fig. \ref{fig:hc_rgb}). Regarding UGC929, we have used deep imaging from DECaLS to study its morphology. Fig. \ref{fig:ugc929} shows the composite DECaLS image of the galaxy. We can see in the image that the galaxy appears to be very symmetric, with no sign of perturbations or interactions. Therefore, if \dg{} and UGC929 had met in the past, it would have been long enough for the central shape of both galaxies to have already been restored.

However, in the outermost parts of \dg{} it would be possible to study the effects of this gravitational interaction in the form of tidal distortions. These deformations produce an S-shaped structure \citep[e.g.][]{Johnston2002, Moreno2022}. Thanks to the deep imaging provided by \hipercam{}, we can investigate whether there is any evidence that this could be the case for \dg. An interaction capable of distorting the galaxy in this way would be clearly visible in our ultra-deep imaging ($\sim$30.5 mag/arcsec$^2$; see Sec. 2.1) even long after the interaction \citep[][]{Moreno2022}. Fig. \ref{fig:hc_rgb} shows an RGB composite of the image of \dg{} with a black and white background to highlight the fainter outskirts of this galaxy. A visual inspection of \dg{} shows no obvious signatures of deformation in the outer regions. This can be seen from the lack of an excess in the outer parts of the surface brightness and stellar density profiles (Fig. \ref{fig:profiles} and Fig. \ref{fig:color}), which is associated with the presence of tidal tails \citep{Johnston2002, Montes2020}.

\begin{figure*}
 \centering
   \includegraphics[width = 0.85\textwidth]{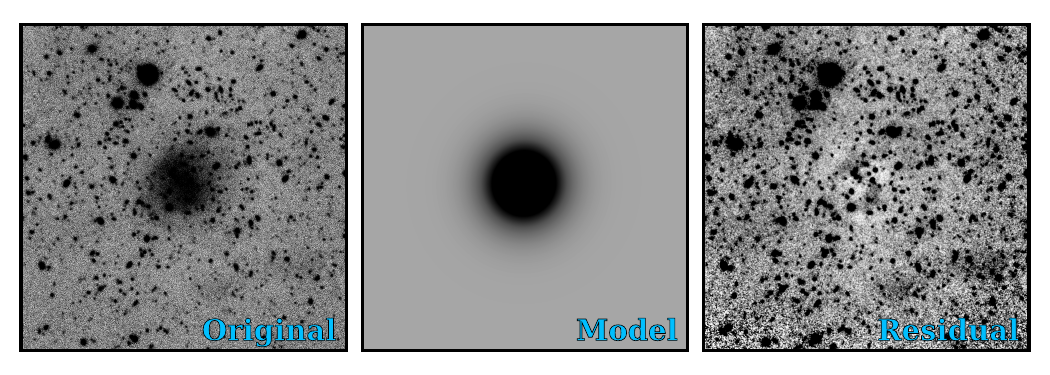}
   \caption{GALFIT model of \dg{} for the \hipercam{} $g+r$-band. The left panel shows the g+r original image, the middle panel the 2-D S\'ersic model fit and the right panel the residuals after subtracting the S\'ersic model, after smoothing it with a Gaussian kernel to enhance the residuals. \label{fig:galfit}}
    \end{figure*}

To improve the visualization of the outer parts of \dg{}, we also modelled \dg{} with GALFIT. To obtain additional signal-to-noise, we have performed a model fit on a $g+r$ image. Fig. \ref{fig:galfit} shows the original (left), model (middle) and residual (right) $g+r$ image. No signs of an excess or deformation can be seen.

\begin{table*}[t]
\centering
\tabcolsep=0.2cm
\begin{tabular}{cccccccc}
\hline \hline
filter & $r_e$ & n  & $b/a$ \\
     &(arcsec) &  & \\
\hline
$u$ & $15.6 \pm 0.6$ & $0.60 \pm 0.03$ & 0.999 \\
$g$ & $15.9 \pm 0.1$ & $0.70 \pm 0.01$ & 0.986 \\
$r$ & $15.3 \pm 0.1$ & $0.65 \pm 0.01$ & 0.962 \\
$i$ & $13.9 \pm 0.1$ & $0.60 \pm 0.01$ & 0.975 \\
$z$ & $16.0 \pm 0.3$ & $0.75 \pm 0.02$ & 0.971 \\
\hline
\end{tabular}
\caption{Multi-band effective radii, axis ration and S\'ersic indexes of the GALFIT \citep{Peng2002} fits to \dg.}\label{app:galfit}
\label{table:galfit}
\end{table*}

\section{\dg{} as a test-bed to explore the nature of dark matter}

\subsection{\dg{} within the cold dark matter scenario}
Given the extreme properties of \dg{}, it is interesting to see whether traditional models of cold dark matter are able to reproduce galaxies with these characteristics. In particular, we are interested in knowing if galaxy formation models can produce objects with stellar masses, surface brightnesses and effective radii like \dg{}. To answer this question, it is worth looking at simulations that have been able to reproduce the properties of the largest known ultra-diffuse galaxies. These simulations have sufficient spatial resolution and stellar population feedback recipes to produce low surface brightness, large effective radius galaxies like DF44.

Using the FIRE-2 simulations \citep{Hopkins2018}, \citet{Chan2018} find that isolated field dwarfs, where the effect of stellar feedback (i.e., stellar winds, supernovae, etc.) is large, produce galaxies with surface brightness, effective radius, and stellar masses representative of UDGs (after imposing artificial quenching to simulate the effect of infall into the cluster environment). Simulated galaxies with stellar masses around 10$^8$ M$_\odot$ and dark matter halos around 10$^{10}$ M$_\odot$, such as \dg{}, have effective radii less than 5 kpc, smaller than what we measure for \dg{}. NIHAO simulations \citep{Wang2015} produce galaxies with similar structural properties to UDGs due to episodes of gas outflows associated with star formation \citep{DiCintio2017}. The dark matter halo masses and stellar masses are consistent with what we find for \dg{}. However, the effective radii ($\sim$3 kpc) of these simulated galaxies are again well below the value we measure for \dg{}.

Other ways that have been proposed to form UDGs are high-z major mergers \citep{Wright2021} or that UDGs populate higher spin halos \citep{Amorisco2016, Benavides2023}. However, neither of these scenarios can reproduce the observed characteristics of \dg. In \citet{Wright2021}, major mergers produce galaxies with $r_e<4$ kpc. \citet{Benavides2023} explores UDGs in the TNG50 simulation and finds that for isolated galaxies of the mass of \dg{}, the effective radius is $\lesssim$ 5 kpc.

\subsection{\dg{} in the fuzzy dark matter framework}

As the microphysical nature of the dark matter is still completely unknown, it is worth exploring whether extreme objects such as \dg{} can be compatible with other alternative dark matter models. In particular, models based on ultralight (axion-like) scalar particles are gaining recently a significant interest \citep[fuzzy dark matter; ][]{Schive2014,2021MNRAS.506.2603M}. Due to the very small mass of these particles ($\sim$10$^{-22}$ eV), the quantum effects are expected to appear at the kpc (i.e., galactic) scale. The ultralight particles generate dark matter models with large cores. In particular, the dark matter distribution is expected to generate a central distribution named soliton followed by a NFW profile in the outer parts \citep{Schive2014}. The soliton density profile can be well approximated by:

\begin{equation}
    \rho_s(r)=\frac{1.9(m_B/10^{-23} eV)^{-2}(r_c/kpc)^{-4}}{[1+9.1\times10^{-2}(r/r_c)^2]^8} M{_\odot} pc^{-3}
\end{equation}

with m$_B$ the mass of the dark matter particle, r$_c$ the core radius where the density has dropped to one-half its peak value. Equation 29 of \citet{Bar2018} shows that if the core radius and the virial mass of the dark matter halo M$_h$ is known, the dark matter particle can be estimated through the following equation:

\begin{equation}
    \frac{m_B}{10^{-22} eV} = 160 \left(\frac{r_c}{pc}\right)^{-1} \left(\frac{M_h}{10^{12} M{_\odot}}\right)^{-1/3}
\end{equation}

To get a rough estimate of the core radius, we assume the stellar distribution follows the shape of the dark matter soliton. This is a reasonable assumption, as the dark matter is the dominant component generating the global gravitational potential \citep{Sanchez-Almeida2023}. Therefore, we project the soliton density profile obtaining:

\begin{equation}
   \Sigma_\star(R) = \frac{\Sigma_\star(0)}{(1 + 0.091\times(R/r_c)^2)^{15/2}} 
\end{equation}

\begin{figure}
 \centering
   \includegraphics[width = 0.45\textwidth]{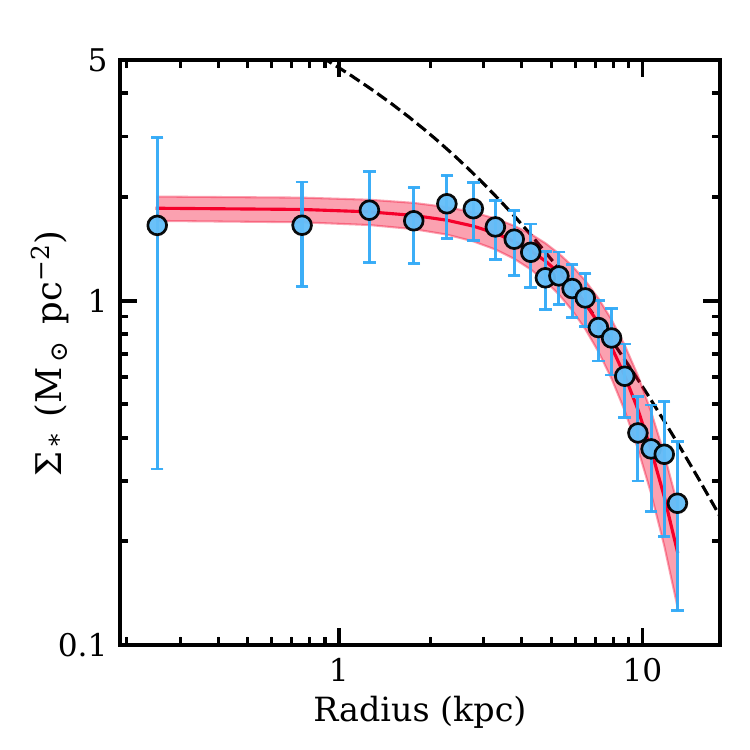}
   \caption{Surface stellar mass density profile of \dg{} (blue dots) fitted with a projected soliton shape with core radius of $r_c$ = $6.6 \pm 0.4$ kpc (see text for details). The soliton best fit and its errors are indicated by the red line and the red region. The agreement is remarkable taking into account that the only free parameter is essentially the core radius. The black dashed line is a projected NFW profile for comparison.\label{fig:densfit}}
    \end{figure}

The stellar mass surface density of \dg{} fits very well with this equation (see Fig. \ref{fig:densfit}). We obtain a value of $r_c$ = 6.6 $\pm$ 0.4 kpc. The mass of the dark matter particle compatible with this core radius is m$_B$=($0.8^{+0.4}_{-0.2}$)$ \times$10$^{-23}$ eV.

We also plot a Navarro, Frenk \& White profile \citep[dashed black line,][]{NFW1996} to show that this commonly used profile to describe the distribution of mass in dark matter halos cannot reproduce the characteristics of \dg{} especially at small radii. For Fig. \ref{fig:densfit}, we used the projected NFW profile in equation 11 in \citet{Wright2000}, assuming as transition radius ($r_s$) the core radius of the soliton derived above. 

It is interesting to compare the mass of the axion-like particle derived in this analysis with other studies. Using galactic rotation curves, \citet{Bernal2018} found an average value of m$_B$$\sim$0.5$\times$10$^{-23}$ eV, while \citet{Banares-Hernandez2023} derived m$_B$$\sim$2$\times$10$^{-23}$ eV. These values are in good agreement with our estimate using only the structural properties of the galaxy. Analysing the velocity dispersion of the Fornax and Sculptor dwarfs, \citet{Gonzalez-Morales2017} find an upper limit of m$_B$$<$4$\times$10$^{-23}$ eV. Based on the angular scale of the CMB acoustic peaks and anisotropies, \citet{Hlovek2018} derive a lower limit of m$_B$$>$0.1$\times$10$^{-23}$ eV. Analysis of the stellar heating of the Milky Way disc due to the substructure within a fuzzy dark matter halo suggests an upper limit of m$_B$$<$4$\times$10$^{-23}$ eV \citep{Chiang2023}. These results are consistent with our estimate for the mass of the axion-like particle. 

However, the situation is far from clear. Other estimates of the mass of the particle are inconsistent with the above results. To name a few, \citet{Chen2017} using Jeans analysis of the Milky Way satellite dSphs find m$_B$=(17.9$\pm$3.3)$\times$10$^{-23}$ eV. \citet{Dalal2022} suggest that the size and stellar kinematics of ultra-faint dwarfs imply a lower limit of m$_B$$>$3$\times$10$^{-19}$ eV. Although fuzzy dark matter could relieve some of the small scale tensions appearing in the cold dark matter scenario, more work is needed to assess this model. 

\section{Conclusions}

In the absence of a laboratory detection of the particles that make up dark matter, the hope for characterising its nature lies in the analysis of astrophysical objects. In particular, galaxies with very low surface brightness (also called ``almost dark'') offer an interesting possibility to constrain the microphysical properties of dark matter. In these galaxies, baryonic feedback effects are expected to be very moderate, leaving the spatial distribution of their dark matter halo almost intact.

In this paper, we present the serendipitous discovery in the IAC Stripe 82 Legacy Project \citep{Fliri2016, Roman2018_S82} of a galaxy, \dg{}, of stellar mass M$_*$ = $3.9\pm1.0\times10^8$ M$_{\odot}$, very extended ($R_e=6.9\pm0.8$ kpc) and very low surface brightness ($<$$\mu_V$$>_e=26.75\pm0.02$ mag/arcsec$^2$) located at 107 Mpc. Using the GBT radio telescope, we inferred a total halo mass of $2.6\pm 1.7 \times10^{10}$ M$_{\odot}$ for the galaxy. Current simulations of ultra-diffuse galaxy formation, which take into account baryonic feedback effects and simulate dark matter particles as WIMPs (i.e., cold dark matter), are unable to reproduce objects with the properties of \dg{}. We have investigated the possibility that the object could be reproduced with the predictions for the fuzzy dark matter model. To this end, and under the hypothesis that the distribution of stars in \dg{} is representative of the distribution of the dark matter halo, we found that a soliton-shaped profile (typical of fuzzy dark matter) reproduces the observed distribution of stars very well. The mass of the axion-like particle inferred from the fit is m$_B$=($0.8^{+0.4}_{-0.2}$)$ \times$10$^{-23}$ eV. This value is in good agreement with other astrophysical measurements using the dynamical properties of other low surface brightness galaxies.

\begin{acknowledgements}
We thank the referee for their useful comments that helped improve the original manuscript. We are indebted to Amelia Trujillo Gonz\'alez for suggesting the name of the galaxy explored in this work. The authors want to thank Scott Carlsten for providing the tables of the structural parameters of the dwarf galaxies in \citet{Carlsten2021}. We also thank Betsy Adams and Pierre-Alain Duc for useful discussions.
This publication is part of the Project PCI2021-122072-2B, financed by MICIN/AEI/10.13039/501100011033, and the European Union “NextGenerationEU”/RTRP.
I.T. acknowledges support from the ACIISI, Consejer\'{i}a de Econom\'{i}a, Conocimiento y Empleo del Gobierno de Canarias and the European Regional Development Fund (ERDF) under grant with reference PROID2021010044 and from the State Research Agency (AEI-MCINN) of the Spanish Ministry of Science and Innovation under the grant PID2019-107427GB-C32, financed by the Ministry of Science and Innovation, through the State Budget and by the Canary Islands Department of Economy, Knowledge and Employment, through the Regional Budget of the Autonomous Community. MM and IT acknowledge support from IAC project P/302302. NC acknowledges support from the research project grant ‘Understanding the Dynamic Universe’ funded by the Knut and Alice Wallenberg Foundation under Dnr KAW 2018.0067.

Based on observations made with the GTC telescope, in the Spanish Observatorio del Roque de los Muchachos of the Instituto de Astrof\'isica de Canarias, under Director’s Discretionary Time.

This work makes use of the following code:
\texttt{astropy} \citep{Astropy2018},  
          \sextractor{} \citep{Bertin1996},
          \scamp{} \citep{Bertin2006},
          \swarp{} \citep{Bertin2010},
          \texttt{Gnuastro} \citep{Akhlaghi2015},     
          \texttt{photutils} v0.7.2 \citep{Bradley2019},
          \texttt{pillow} \citep{pillow2020},
          \texttt{numpy} \citep{oliphant2006},
          \texttt{scipy} \citep{scipy2020},
          \texttt{Astrometry.net} \citep{Lang2010},
          \texttt{GALFIT} \citep{Peng2002}.
\end{acknowledgements}

\bibliography{dgs82.bib}
\bibliographystyle{aa}

\appendix

\section{Properties of UGC929} \label{app:ugc929}

Given the difference in distance between both galaxies, we assumed that \dg{} is a satellite of UGC 929. UGC 929 is at a projected distance of 14.7$\arcmin$ (435 kpc at a distance of 107 Mpc) from \dg{}. To calculate the stellar mass of UGC929, we use the DECaLS g and r images, obtaining the M/L ratio using the \citet{Roediger2015} prescriptions and assuming a \citet{Chabrier2003}. The stellar mass of the galaxy is $M_* = 2.4\times10^{10}M_\odot$. 

 \begin{figure}
 \centering
  \includegraphics[height = 0.35\textheight]{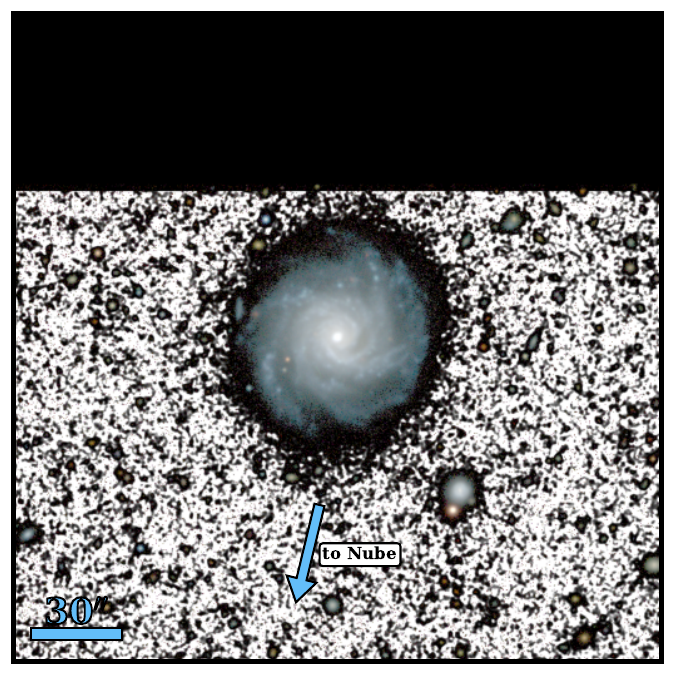}
  \caption{ A region of $250\arcsec\times250\arcsec$ ($\sim$123 kpc $\times$ 123 kpc) around the galaxy UGC929. The figure is a composite of an RGB image using the $g$, $r$ and $z$ from DECaLS and a black and white $g+r+z$ image for the background.
  This galaxy is at the same redshift that \dg, but 14.7\arcmin\ to the north-west. The blue arrow indicates the direction of \dg. There is no evidence of morphological disturbance that could point to gas expulsion as the origin of \dg. 
  \label{fig:ugc929}}
    \end{figure}

\section{Alternative distances to \dg}\label{app:distance}

Some of the most peculiar properties of \dg{} depend on the assumed distance to the galaxy. In Section \ref{sec:hIcont} we found that it is unlikely that another nearby galaxy could be responsible for the HI emission we detected. However, we cannot completely rule out the possibility that the HI detection is not related to \dg{}. In that case, the galaxy could be at a different distance from the one assumed in this work. For this reason, it is worth exploring what alternative distances are potentially possible.

\begin{figure*}
 \centering
   \includegraphics[height = 0.55\textheight]{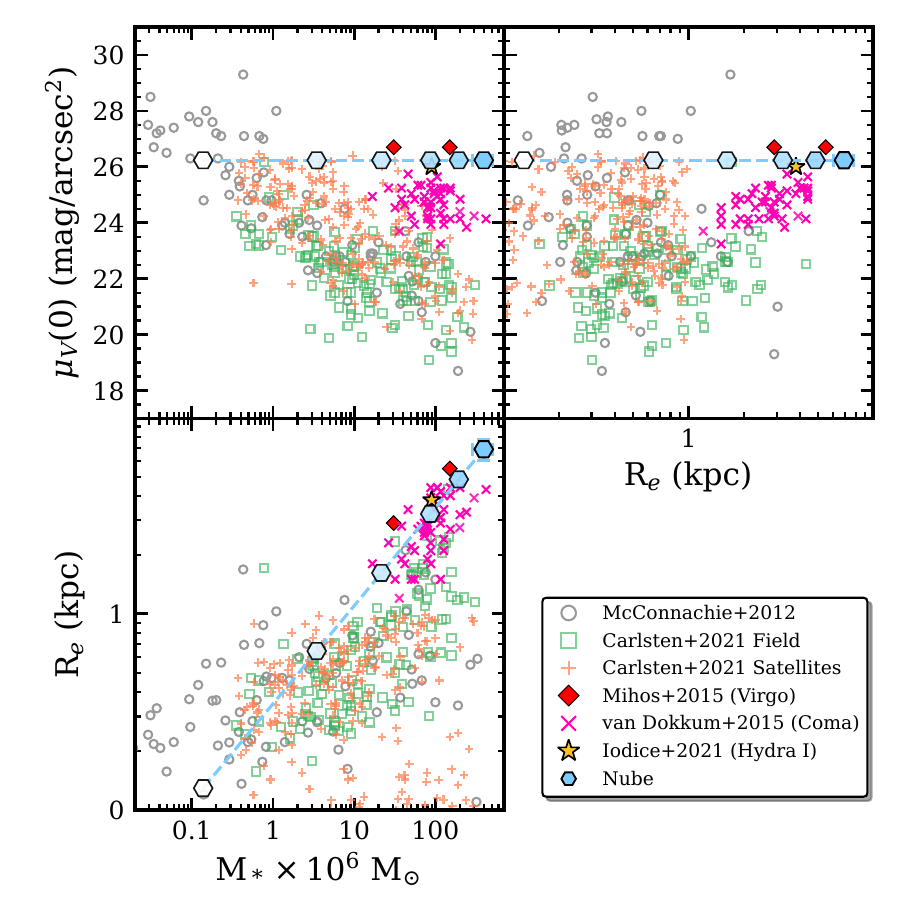}
   \caption{Same as Fig. \ref{fig:comparison} but showing \dg{} at different distances 2, 10, 25, 50, 75 and 107 Mpc. }\label{fig:comp_dist}
    \end{figure*}

One possibility is that \dg{} is a satellite of a galaxy in the field of view other than UGC929. Exploring the nearby galaxies with spectroscopic redshifts (Fig. \ref{fig:fov}), \dg{} could be associated with UGC928 (8248 km/s). However, if it is associated with UGC928 (projected separation of $\sim3.8\arcmin$), \dg{} will be even further away and its properties will be even more extreme. Alternatively, \dg{} could be closer to us. In Fig. \ref{fig:comp_dist} we show the inferred properties of \dg{} at different distances, from 2 Mpc (white) to the assumed distance of 107 Mpc (darker blue). The only galaxy in the field of view with a spectroscopic redshift that is closer to us is UGC931 (Fig. \ref{fig:fov}) at $\sim30$ Mpc \citep{Springob2009}. If \dg{} it is associated with UGC931, its properties will also be peculiar. Its stellar mass and $R_e$ will be similar to those of UDGs, but it would have a significantly lower central surface brightness (3rd to 4th hexagon from the left, upper right panel in Fig. \ref{fig:comp_dist}). 

For \dg{} to be a more normal galaxy, the distance should be between 2 and 10 Mpc. In this case, \dg{} will be more like the low-mass Local Group galaxies of \citet{McConnachie2012}, such as \textsc{Phoenix} (M$_*$ = $0.7\times10^6$ M$_\odot$, $r_e$ = 454 pc and $\mu_V(0)$ = 25.8 mag/arcsec$^2$) or the dwarfs of \citet{Carlsten2021}. However, there is no nearby massive galaxy to act as a host for \dg{} in such case, and therefore it would be a rare example of an isolated very low-mass galaxy.

\section{Photometric redshifts and other sources around \dg}

Given the spatial resolution of the GBT data, in section \ref{sec:hIcont} we discussed the possibility that the HI detection corresponds to another source and not to \dg{}. Table \ref{table:photz} lists all galaxies with photometric redshifts within 10\arcmin\ of \dg. Table \ref{table:other_cont} lists faint sources that have neither spectroscopic nor photometric redshifts, but could be responsible for the HI emission.
 
\begin{table*}[t]
\centering
\tabcolsep=0.2cm
\begin{tabular}{cccccccc}
\hline \hline
RA & DEC & z & m$_r$ & distance  \\
 (deg)  &(deg) &  & (AB mag) & (arcmin)\\ \hline
20.7455 & -0.6473 & 0.028$ \pm $0.03 & 17.3 & 7.3*\\
20.8104 & -0.6972 & 0.025$ \pm $0.03 &  20.4  & 5.5$^a$\\
20.8105 & -0.7059 & 0.025$ \pm $0.02  & 21.0  & 5.9$^a$ \\
20.8349 &  -0.6543 & 0.025$ \pm $0.1  & 20.3  & 2.5* \\
20.7561 &  -0.5262 & 0.027$ \pm $0.05 &  17.2  & 8.9* \\ 
20.7682 &  -0.5045 & 0.026$ \pm $0.2  & 20.2  & 9.2* \\
20.7984  & -0.6008 & 0.025$ \pm $0.1  & 20.6  & 4.2* \\
20.8535  & -0.5496 & 0.026$ \pm $0.1  & 20.0  & 4.5* \\
20.8571 &  -0.5475 & 0.025$ \pm $0.01 & 12.1 &  4.6* \\
20.9294 &  -0.4799 & 0.026$ \pm $0.09 & 20.4  & 9.5* \\
21.0163 &  -0.6003 & 0.027$ \pm $0.15  &  20.5  & 9.2* \\
\hline
\hline
\end{tabular}
\caption{List of objects with photometric redshifts compatible with the HI detection reported in this paper. The vast majority of these objects are stars or point-like sources (marked with *), another two sources (marked with $^a$) are part of a known galaxy, UGC931.}\label{app:photz}
\label{table:photz}
\end{table*}

\begin{table*}[t]
\centering
\tabcolsep=0.2cm
\begin{tabular}{cccccccc}
\hline \hline
RA & DEC & Distance to \dg{} \\
(deg)  & (deg) & (arcmin)  \\ \hline
20.9390 & -0.6132 & 4.5 \\
20.9092 &  -0.5122 & 7.2 \\
\hline
\hline
\end{tabular}
\caption{List of faint sources around \dg{} that we cannot exclude as responsible for the HI emission reported in this work.}\label{app:cont}
\label{table:other_cont}
\end{table*}

\section{Masks for photometry and surface brightness profiles}\label{app:mask}

\begin{figure}
 \centering
   \includegraphics[height = 0.5\textwidth]{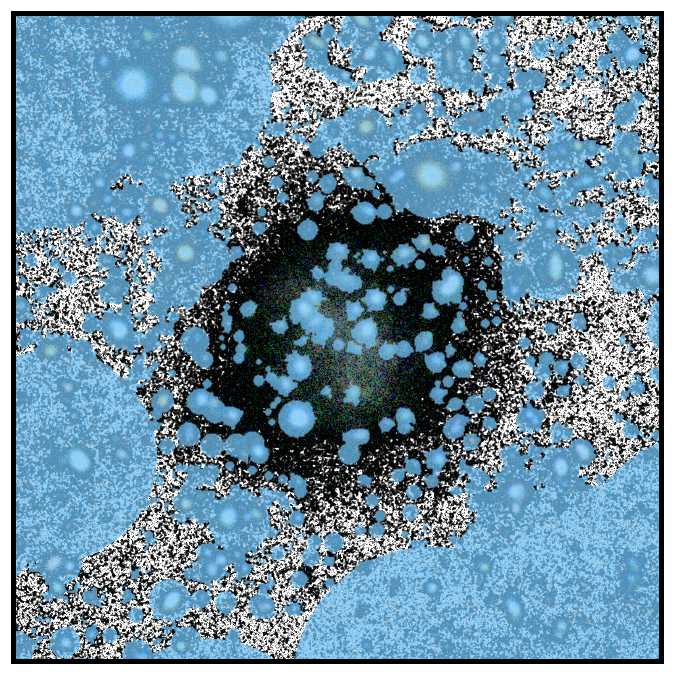}
   \caption{Mask (blue regions) applied to a HiPERCAM RGB colour (g+r+i) image of a region $100\arcsec\times100\arcsec$ around \dg. The black and white background is a $g+r$ image. The image shows the need for thorough masking in these ultra-deep images.}\label{fig:mask}
    \end{figure}

Masking is a key component in obtaining reliable photometry and surface brightness profiles of the galaxy under study. In section \ref{sec:prof} we describe how the HiPERCAM mask used to derive the photometry and surface brightness profiles of \dg was created. Fig. \ref{fig:mask} shows the mask used to derive the photometry and surface brightness profiles of the galaxy. 

\end{document}